\documentclass[12pt]{article}
\input tcilatex
\usepackage{latexsym}

\begin{document}

\author{A. E. Santana${}^{a,b}$, F. C. Khanna${}^{b,c}$ and Y. Takahashi${}^{b}$ \\
${}^{a}$Instituto de Fisica, Universidade Federal da Bahia,\\
Campus de Ondina, 40210-340,\\
Salvador, Bahia, Brazil;\\
${}^b$Theoretical Physics Institute, Dept. of Physics, \\
University of Alberta, \\
Edmonton, AB T6G 2J1, Canada; and\\
${}^c$TRIUMF, 4004, Wesbrook Mall,\\
Vancouver, BC V6T 2A3, Canada}
\title{Galilei Covariance and (4,1)-de Sitter Space}
\maketitle

\begin{abstract}
A vector space $\mathcal{G}$ is introduced such that the Galilei
transformations are considered linear mappings in this manifold. The
covariant structure of the Galilei Group (Y. Takahashi, Fortschr. Phys%
\textbf{. 36} (1988) 63, \textbf{36} (1988) 83) is derived and the tensor
analysis is developed. It is shown that the Euclidean space is embedded the
(4,1) de Sitter space through in $\mathcal{G}$. This is an interesting and
useful aspect, in particular, for the analysis carried out for the Lie
algebra of the generators of linear transformations in $\mathcal{G}$.
\end{abstract}

\newpage
Galilean symmetries constitute a natural scenario for formulations of
non-relativistic theories, with particular emphasis to the study of the
condensed matter physics \cite{g1}-\cite{g6}. Unlike the Poincar\'e group,
however, the representations of the Galilei group ($G$) have not been
developed sufficiently \cite{g5}, even though there is a wealth of
informations about non-Lorentzian physics that could benefit greately from
such studies.

One aspect that makes such a development difficult relies on the intricate
structure of $G$, characterized by eleven parameters: three spatial
rotations, three spatial translations, three boosts, one time translation
and one central extension (necessary in order to find physical
representations). The natural representation, nevertheless, is described by
10 parameters and specified by

\begin{eqnarray}
{\bar {\mathbf{x}}} &=&R\mathbf{x}+\mathbf{v}t+\mathbf{a},  \label{trans1} \\
{\bar t} &=&t+b.  \label{trans2}
\end{eqnarray}

Usually $G$ has been introduced without the metrical vector space in which
the transformations given by Eqs.(\ref{trans1}) and (\ref{trans2}) are
defined, as it is the case of the group $O(3)$, defined on the $\mathcal{R}%
^3 $-space, or the Poincar\'e group, defined as linear transformations in
the Minkowsky space. This lack of a Galilean metric-vector space has a
consequence that a ray representation of $G$ is not trivially reducible to a
faithful representation. Therefore, it is interesting, for the study of the
Galilei symmetries, to specify the manifold underlying the Galilean
transformations. The objective of this paper is to give the Galilean group
also as a linear isometry group, similar to the case, for instance, of the
Poincar\'e group.

A tensor structure for the Galilean transformations was undertaken some time
ago by one of the authors~\cite{takahashi1,takahashi2}. Such a structure is
based in a five dimensional formulation of the Galilei transformations, and
has been motived by the development of the Galilei invariant field theory.
For instance, this formalism has been used to introduce generalized
Schr\"odinger equations, and to derive a non-linear Galilei invariant field
equation, from which the rearrangement of symmetries describing rotons and
phonons has been studied.

Here, developments of that covariant approach to the Galilei group are
presented. In particular, we introduce the tensor formulation, stressing its
manifold characteristics in order to build the manifold analysis. In this
sense, we develop representations of the Galilei Lie algebra on such a
manifold (say $\mathcal{G}$), taking advantage of the (if not intriguing, at
least practical) fact that $\mathcal{G}$ can be considered as an embedding
of the 3-dimensional Euclidean space $(\mathcal{E})$ in a $(4,1)$ de Sitter
space\cite{g8} (another standpoint to introduce this 5-dim structure is
motived by the Newton-Cartan theory of gravitation, see Ref.\cite{kun}).

Considering kinematic groups, the Galilei group has been studied previously
via a Wigner-In{\"o}n{\"o}u contraction of the Poincar\'e group, which is,
in turn, contracted from the de Sitter group \cite{g7}. This, however, is
not the case for the approach developed here, where the concept of
embedding, involving geometrical structures, without any limiting process,
is used. This allows us to establish a direct link between the Galilei and
de Sitter groups.

Let us begin by observing that in $\mathcal{E}$, the metric space defined on 
$\mathcal{R}^3$, the distance between two points is preserved under linear
transformations. That is, given two vectors $\mathbf{x}=(x^1,x^2,x^3)$ and $%
\mathbf{y}=(y^1,y^2,y^3); \mathbf{x},\mathbf{y}\in \mathcal{E}$, then $r^2=%
\mathbf{x}^2+\mathbf{y}^2-2\mathbf{x}\mathbf{y}$ is invariant under
translations and rotations. In a physical system described by Galilei
symmetries, the two types of translations of $G$ occur in $\mathcal{E}$.
However, one of them , the boost, is defined via an external parameter, the
time $t$ (see Eqs.(\ref{trans1}) and (\ref{trans2})); and this is a central
aspect of ${G}$.

We can consider, therefore, using the form of the distance ($r$) in the
Euclidean space, to embed $\mathcal{E}$ in a lager manifold, say $\mathcal{G}
$, such that Eqs.(\ref{trans1}) and (\ref{trans2}) can be considered as
linear transformations in $\mathcal{G}$. This can be achieved, indeed, if we
observe that

\begin{equation}
s^2=-{\frac 12}r^2=-t{\frac{\mathbf{x}^2}{2t}}-t{\frac{\mathbf{y}^2}{2t}}+%
\mathbf{x} \cdot \mathbf{y}  \label{metrica1}
\end{equation}
is no more than the inner product of two particular vectors of a space $%
\mathcal{G}$, which is defined as follows.

Let $\mathcal{G}$ be a 5-dimensional metric space, with an arbitrary vector
denoted by $x=(x^1,x^2,x^3,x^4,x^5)$ = $(\mathbf{x},x^4,x^5)$. The inner
product in $\mathcal{G}$ is defined by 
\begin{eqnarray}
(x|y) &=&{\eta }_{\mu \nu }x^\mu y^\nu   \nonumber \\
&=&\sum_{i=1}^3{x^iy^i}-x^4y^5-x^5y^4,  \label{pescal}
\end{eqnarray}
where $x,y\ \in \mathcal{G}$ and $\eta _{\mu \nu }$, the metric, is given by
(Latin indices represent components of vectors in $\mathcal{E}$, as in Eq.(%
\ref{pescal}))\cite{takahashi1,takahashi2,takahashi3,kun} 
\begin{equation}
({\eta }_{\mu \nu })=\left( 
\begin{array}{ccccc}
1 & 0 & 0 & 0 & 0 \\ 
0 & 1 & 0 & 0 & 0 \\ 
0 & 0 & 1 & 0 & 0 \\ 
0 & 0 & 0 & 0 & -1 \\ 
0 & 0 & 0 & -1 & 0
\end{array}
\right) .  \label{metr1}
\end{equation}
(A similar metric structure was proposed by Cangemi and Jackiw\cite{jac} in
order to treat a $(1+1)$ gravitational theory as a gauge formalism.)

Notice that $s^2$, defined in Eq.(\ref{metrica1}), is a particular case of
the inner product in $\mathcal{G}$. That this is the case can be seen by
writing

\begin{equation}
x^5={\frac{\mathbf{x}^2}{2t}},\quad y^5={\frac{\mathbf{y}^2}{2t}},\quad 
\text{and}\quad x^4=y^4=t,  \label{metrica4}
\end{equation}
in Eq.(\ref{pescal}). (In order to adjust the physical units of space and
time, we can define, for instance, $x^4=vt,$ with $v=1m/s$.)

Let $\{e_\mu \}=\left\{ e_1,...,e_5\right\} $ be a basis vector of $\mathcal{%
G}$, such that $x=x^\mu e_\mu $ and $y=y^\mu e_\mu $. Then, from Eq.(\ref
{pescal}), it follows that $(e_\mu |e_\nu )=\eta _{\mu \nu }=\eta _{\nu \mu
} $. In addition, the dual structure of $\mathcal{G}$ can be introduced.
Consider $\mathcal{G}^{*}=\left\{ w_1,w_2,...\right\} $ the set of linear
forms on $\mathcal{G}$; that is, $w:\mathcal{G}\ \mapsto \ \mathcal{R}$, so
that $(w_1+aw_2)\ x=w_1(x)+aw_2(x),\,\,a\in \mathcal{R}$, and

\begin{equation}
w(x)=x^\mu w(e_\mu )=w_\mu x^\mu ,  \label{forma1}
\end{equation}
where $w_\mu =w(e_\mu )$. $\mathcal{G}^{*}$ is defined by the following set
of 1-forms$\ \ e^\mu :\ x\quad \mapsto x^\mu \ $,  $e^\mu (x)=x^\mu .$
Therefore, from Eq.(\ref{forma1}), we can write $w=w_\mu e^\mu $, so that $%
\left\{ e^\mu \right\} $ is a dual basis of $\mathcal{G}$.

The metric $\ \eta _{\mu \nu }$ can be used to define properly the operation
of raising and lowering of indices. In order to do so, first, one uses Eq.(%
\ref{pescal}) to introduce a natural 1-form (also called natural pairing~%
\cite{bishop}) defined by $x^{*}(y)\equiv (x|y).$ Second, writing $x=x^\mu
e_\mu $ and $x^{*}=x_\nu e^\nu $, it follows that $x^{*}(y)=x_\mu y^\nu
e^\mu (e_\nu );$ and it leads to $e^\mu (e_\nu )=\delta _{\ \nu }^\mu $.
Considering then the definition of $x^{*}$ and the fact that $y$ is an
arbitrary vector, the operation of lowering indices is established, that is $%
x_\mu =\eta _{\mu \nu}x^\nu. $ Introducing $(\eta ^{\mu \nu })^{-1}=(\eta
_{\mu \nu })$, the operation of raising indices is $x^\mu =\eta ^{\mu \nu
}x_\nu .$ As a result, Eq.(\ref{pescal}) can be written as $(x|y)=x^\mu
y_\mu =x_\mu y^\mu ,$ such that 
\begin{eqnarray}
e^i(x) &=&x^i=x_i,\quad i=1,2,3,  \nonumber \\
e^4(x) &=&x^4=-x_5,  \nonumber \\
e^5(x) &=&x^5=-x_4.  \nonumber
\end{eqnarray}

The norm of a vector in $\mathcal{G}$ is defined as $|\!|x|\!|=(\mathbf{x}%
)^2+x_4x^4+x_5x^5=(\mathbf{x})^2-2x^4x^5.$ If $|\!|x|\!|>0$ and $x^4$ and $%
x^5$ are real numbers with the same sign, then $\mathbf{x}^2\neq 0$. In this
case, following the Minkowsky space example, $x$ will be called a space-like
vector. Null-like vectors are those with $|\!|x|\!|=0$, that is, $(\mathbf{x}%
)^2=2x^4x^5$. Therefore, the condition $|\!|x|\!|\geq 0$ is physically
acceptable, since the movement of the system is in a manifold with the space
in both cases given by $(\mathbf{x})^2\geq 0$. For vectors of type $%
|\!|x|\!|<0$, the physically acceptable situations are those for which $x^4$
and $x^5$ have the same sign, for $(\mathbf{x})^2<2x^4x^5$.

Each vector in $\mathcal{E}$, say $\mathbf{A}=(A^1,A^2,A^3)$, is in a
correspondence with a vector in $\mathcal{G}$, say $A$, through the
embedding, $\Im :\mathbf{A}\ \mapsto \ A=(\mathbf{A},d,{\mathbf{A}^2/2d})$,
where $d$ is an arbitrary quantity. Indeed, using Eq.(\ref{pescal}), it
follows that, in this case,

\begin{eqnarray*}
(A|A) &=&{\eta }_{\mu \nu }A^\mu A^\nu , \\
&=&\sum_{i=1}^{3}{A^iA^i}-2A^4A^5=0.
\end{eqnarray*}
That is, according to $\Im $, each vector in $\mathcal{E}$ is in a
homomorphic correspondence with null-like vectors in $\mathcal{G}$.

$\mathcal{G}$ can still be mapped into a $(4,1)$-de Sitter space $(\mathcal{S%
})$\cite{g8} by the following linear transformation, $U$, \cite{takahashi3} 
\begin{eqnarray}
U &:&x^i\mapsto \xi ^i=x^i,\,\,\,i=1,2,3,  \label{t1} \\
U &:&x^4\mapsto \xi ^4=(x^4+x^5)/\sqrt{2}, \\
U &:&x^5\mapsto \xi ^5=(x^4-x^5)/\sqrt{2};  \label{t2}
\end{eqnarray}
resulting in 
\begin{equation}  \label{onze}
(x|y)=g_{\mu \nu }\xi^\mu \zeta^\nu=(\xi |\zeta ),
\end{equation}
with the (diagonal) metric tensor ($g_{\mu \nu }$) specified by $%
diag\,(g_{\mu \nu })=(+,+,+,-,+)$ (general vectors in $\mathcal{S}$ are
being denoted by Greek letters as $\xi $, $\zeta $, $\varsigma$, and so on )
. In short, we can gather the above results stating: 
\[
\]
\textbf{Proposition}: Using the $\mathcal{G}$ manifold, $\mathcal{E}$ can be
embedded into $\mathcal{S}$, a de Sitter space, through the composit mapping 
$U{\circ }\Im :\mathcal{E\mapsto S}$, where the transformation $U$ is given
by 
\begin{equation}
U=(U_{\,\,\nu }^\mu )=\left( 
\begin{array}{ccccc}
1 & 0 & 0 & 0 & 0 \\ 
0 & 1 & 0 & 0 & 0 \\ 
0 & 0 & 1 & 0 & 0 \\ 
0 & 0 & 0 & \frac 1{\sqrt{2}} & \frac 1{\sqrt{2}} \\ 
0 & 0 & 0 & \frac 1{\sqrt{2}} & \frac{-1}{\sqrt{2}}
\end{array}
\right) ,  \label{matrix1}
\end{equation}
such that the mapping $U:x^\mu \mapsto \xi ^\mu $ , $\xi ^\mu \in \mathcal{S}
$, $x ^\mu \in \mathcal{G}$, is given by 
\begin{equation}
\xi ^\mu =U_{\,\,\nu }^\mu x^\nu ,  \label{t3}
\end{equation}
with $U=U^{-1}$. So, in general, an embedded vector $A$ in $\mathcal{S}$
(from the vector $\mathbf{A}$ in $\mathcal{E}$) is given by 
\[
A=(\mathbf{A}, {\frac{1 }{d\sqrt{2} }}(2d+\mathbf{A}^2), {\frac{1 }{d\sqrt{2}
}}(2d-\mathbf{A}^2)).\ \ \ \ \ \ \ \Box 
\]
\[
\]

The transformation matrix $U$ can be used to relate the metric $g$ of $%
\mathcal{S}$ to $\eta$ of $\mathcal{G}$. In fact, according to Eq.(\ref{onze}%
) $(\xi |\zeta )=g_{\mu \nu }\xi ^\mu \zeta ^\nu $; so we get from Eq.(\ref
{t3}) 
\begin{equation}
(\xi |\zeta )=U_{\,\,\rho }^\mu g_{\mu \nu }U_{\,\,\gamma }^\nu x^\rho
x^\gamma .  \label{t4}
\end{equation}
Using then Eq.(\ref{pescal}), we have 
\begin{equation}
\eta _{\rho \gamma }=U_{\,\,\rho }^\mu g_{\mu \nu }U_{\,\,\gamma }^\nu ;
\label{t5}
\end{equation}
or its inverse, $g=U\eta U.$

It is worth observing that we can define another, more restricted, embedding
in $\mathcal{G}$ space by $\Im^{\prime}:\mathbf{A}\ \mapsto \ A=(\mathbf{A}%
,e,0)$, where $e$ is an arbitrary quantity. In this case, on the other hand, 
$A$ is no longer a null-like vector, for $(A|A)=\mathbf{A}^2$. In $\mathcal{S%
}$ space, such a vector is written as $A=(\mathbf{A},e/2,e/2)$.

Simple examples of the two kinds of embedded vectors in $\mathcal{G}$ are
provided by 
\begin{eqnarray}
P&=&(\mathbf{P}, E, m),  \label{mom1} \\
x&=&(x,t,0);
\end{eqnarray}
where, in the former $E=\mathbf{P}^2/2m$ is the energy and $d=m$; while in
the latter $e=t$.

Now we are going to explore linear transformations in the space $\mathcal{G}$%
. Let

\begin{equation}
\overline{{x}}^\mu =G_{\ \nu }^\mu x^\nu ,  \label{transf1}
\end{equation}
be a homogeneous linear transformation, such that the metric tensor $\eta
_{\mu \nu }$ and the inner product, Eq.(\ref{pescal}), are invariant. Then 
\begin{equation}
G\eta G^T=\eta ,  \label{transf2}
\end{equation}
where $G^T$ is the transposed matrix of $G$.

Consider infinitesimal transformations of the connected part of $G$, i.e.,
with $G_{\ \nu }^\mu =\delta _{\ \nu }^\mu +\epsilon _{\ \nu }^\mu $, with $%
|G|=1$. Using Eq.(\ref{transf2}) we obtain 
\begin{equation}  \label{cond1}
\epsilon^{\alpha}_{\ \nu}\eta_{\alpha\beta} +
\eta_{\nu\alpha}\epsilon^{\alpha}_{\ \beta} = 0.
\end{equation}
From the analysis of Eq.(\ref{cond1}), the matrix $(\epsilon^{\mu}_{\ \nu}) $
can be written as 
\begin{equation}
({\epsilon }_{\ \nu }^\mu )=\left( 
\begin{array}{ccccc}
0 & \epsilon _{\,\,2}^1 & \epsilon _{\,\,3}^1 & \epsilon _{\,\,4}^1 & 
\epsilon _{\,\,5}^1 \\ 
-\epsilon _{\,\,2}^1 & 0 & \epsilon _{\,\,3}^2 & \epsilon _{\,\,4}^2 & 
\epsilon _{\,\,5}^2 \\ 
-\epsilon _{\,\,3}^1 & -\epsilon _{\,\,3}^2 & 0 & \epsilon _{\,\,4}^3 & 
\epsilon _{\,\,5}^3 \\ 
\epsilon _{\,\,5}^1 & \epsilon _{\,\,5}^2 & \epsilon _{\,\,5}^3 & \epsilon
_{\,\,4}^4 & 0 \\ 
\epsilon _{\,\,4}^1\  & \epsilon _{\,\,4}^2\  & \epsilon _{\,\,4}^3 & 0 & 
-\epsilon _{\,\,4}^4
\end{array}
\right) .  \label{matrix2}
\end{equation}
Defining 
\begin{eqnarray*}
\epsilon _{\,\,2}^1 &=&m^3,\,\,\epsilon _{\,\,3}^1=m^2,\,\,\epsilon
_{\,\,3}^2=m^1, \\
\epsilon _{\,\,4}^1 &=&n^1,\,\,\,\epsilon _{\,\,4}^2=n^2,\,\,\,\,\epsilon
_{\,\,4}^3=n^3, \\
\epsilon _{\,\,5}^1 &=&u^1,\,\,\,\epsilon _{\,\,5}^2=u^2,\,\,\,\,\epsilon
_{\,\,5}^3=u^3, \\
\epsilon _{\,\,4}^4 &=&u,
\end{eqnarray*}
matrix $({\epsilon }_{\ \nu }^\mu )$, Eq.(\ref{matrix2}), can be written as

\begin{equation}
({\epsilon }_{\ \nu }^\mu )=\sum_{i=1}^{3}{m^iL_i}+\sum_{i=1}^{3}{n^iB_i}%
+\sum_{i=1}^{3}u{^iC_i}+uD,  \label{matrix4}
\end{equation}
where 
\[
L_1=\left( 
\begin{array}{ccccc}
0 & 0 & 0 & 0 & 0 \\ 
0 & 0 & 1 & 0 & 0 \\ 
0 & -1 & 0 & 0 & 0 \\ 
0 & 0 & 0 & 0 & 0 \\ 
0 & 0 & 0 & 0 & 0
\end{array}
\right) ,\,\,\,L_2=\left( 
\begin{array}{ccccc}
0 & 0 & 1 & 0 & 0 \\ 
0 & 0 & 0 & 0 & 0 \\ 
-1 & 0 & 0 & 0 & 0 \\ 
0 & 0 & 0 & 0 & 0 \\ 
0 & 0 & 0 & 0 & 0
\end{array}
\right) 
\]
\[
\,L_3=\left( 
\begin{array}{ccccc}
0 & 1 & 0 & 0 & 0 \\ 
-1 & 0 & 0 & 0 & 0 \\ 
0 & 0 & 0 & 0 & 0 \\ 
0 & 0 & 0 & 0 & 0 \\ 
0 & 0 & 0 & 0 & 0
\end{array}
\right) ,\,\,B_1=\left( 
\begin{array}{ccccc}
0 & 0 & 0 & 1 & 0 \\ 
0 & 0 & 0 & 0 & 0 \\ 
0 & 0 & 0 & 0 & 0 \\ 
0 & 0 & 0 & 0 & 0 \\ 
1 & 0 & 0 & 0 & 0
\end{array}
\right) \,\,\,\,\,\,\,\,\,\,\,\,\,\,\,\,\,\, 
\]
\[
B_2=\left( 
\begin{array}{ccccc}
0 & 0 & 0 & 0 & 0 \\ 
0 & 0 & 0 & 1 & 0 \\ 
0 & 0 & 0 & 0 & 0 \\ 
0 & 0 & 0 & 0 & 0 \\ 
0 & 1 & 0 & 0 & 0
\end{array}
\right) ,\,\,\,B_3=\left( 
\begin{array}{ccccc}
0 & 0 & 0 & 0 & 0 \\ 
0 & 0 & 0 & 0 & 0 \\ 
0 & 0 & 0 & 1 & 0 \\ 
0 & 0 & 0 & 0 & 0 \\ 
0 & 0 & 1 & 0 & 0
\end{array}
\right) \,\,\,\,\,\,\,\,\,\,\,\,\, 
\]
\[
C_1=\left( 
\begin{array}{ccccc}
0 & 0 & 0 & 0 & 1 \\ 
0 & 0 & 0 & 0 & 0 \\ 
0 & 0 & 0 & 0 & 0 \\ 
1 & 0 & 0 & 0 & 0 \\ 
0 & 0 & 0 & 0 & 0
\end{array}
\right) ,\,\,\,C_2=\left( 
\begin{array}{ccccc}
0 & 0 & 0 & 0 & 0 \\ 
0 & 0 & 0 & 0 & 1 \\ 
0 & 0 & 0 & 0 & 0 \\ 
0 & 1 & 0 & 0 & 0 \\ 
0 & 0 & 0 & 0 & 0
\end{array}
\right) \,\,\,\,\, 
\]
\[
C_3=\left( 
\begin{array}{ccccc}
0 & 0 & 0 & 0 & 0 \\ 
0 & 0 & 0 & 0 & 0 \\ 
0 & 0 & 0 & 0 & 1 \\ 
0 & 0 & 1 & 0 & 0 \\ 
0 & 0 & 0 & 0 & 0
\end{array}
\right) ,\,\,\,\,\,\,\,\,D=\left( 
\begin{array}{ccccc}
0 & 0 & 0 & 0 & 0 \\ 
0 & 0 & 0 & 0 & 0 \\ 
0 & 0 & 0 & 0 & 0 \\ 
0 & 0 & 0 & 1 & 0 \\ 
0 & 0 & 0 & 0 & -1
\end{array}
\right) . 
\]

The commutation relations among these generators, $L_1$,..., $D$, give raise
to the following algebraic relations 
\begin{eqnarray}
\lbrack L_i,L_j] &=&\varepsilon _{ijk}L_k,\,\,\,[L_i,C_j]=\varepsilon
_{ijk}C_k,\,\,\,[L_i,B_j]=\varepsilon _{ijk}B_k,  \nonumber \\
\lbrack B_i,C_j] &=&\varepsilon _{ijk}L_k-D\delta
_{ij},\,\,\,[B_i,D]=B_i\,,\,[D,C_i]=C_i.  \label{sutt1}
\end{eqnarray}

In order to study representations of such a Lie algebra, we can take
advantage of the fact that these generators are connected with those of
linear transformations in the de Sitter space, $\mathcal{S}$, since the de
Sitter coordinates, $\xi ^\mu $, are connected to those of the $\mathcal{G}$%
-space, $x^\mu $, by Eq.(\ref{t3}). Then, a linear transformation in the $%
\mathcal{G}$-space, characterized by $G_{\ \nu }^\mu =\delta _{\ \nu }^\mu
+\epsilon _{\ \nu }^\mu $, induces a transformation $\widetilde{G}_{\ \nu
}^\mu \,\,$in $\mathcal{S}$ specified by 
\[
\widetilde{G}_{\ \nu }^\mu =U_{\ \alpha }^\mu G_{\ \beta }^\alpha U_{\ \nu
}^\beta , 
\]
where $S$ is given by Eq.(\ref{matrix1}). To proceed further, let us write
the algebra given by Eqs.(\ref{sutt1}) in a covariant form, i.e., 
\begin{equation}
\lbrack M_{\alpha \beta },M_{\gamma \rho }]=i(\eta_{\alpha \rho }M_{\beta
\gamma }-\eta_{\alpha \gamma }M_{\beta \rho }+\eta_{\beta \gamma }M_{\alpha
\rho }-\eta_{\beta \rho }M_{\alpha \gamma });  \label{sit1}
\end{equation}
with $\alpha ,\beta ,...=1,...,5$, such that 
\begin{eqnarray*}
iM_{ij} &=&\varepsilon_{ijk}L_k,\, \\
iM_{i4} &=&M_{4i}=B_i, \\
iM_{i5} &=&M_{5i}=C_i, \\
iM_{45} &=&M_{45}=D.
\end{eqnarray*}
Another representation for these operators is 
\[
M_{\alpha \beta }=-i(x_\alpha \frac \partial {\partial x_\beta }-x_\beta
\frac \partial {\partial x_\alpha }) 
\]

Using the transformation $U$, the de Sitter Lie algebra and its Casimir
invariants are derived. That means, if we define ${\widetilde{M}}=UMU\ \ $,
with ${\widetilde{M}}\in \mathcal{S}$, we get 
\begin{equation}
\lbrack {\widetilde{M}}_{\alpha \beta },{\widetilde{M}}_{\gamma \rho
}]=i(g_{\alpha \rho }{\widetilde{M}}_{\beta \gamma }-g_{\alpha \gamma }{%
\widetilde{M}}_{\beta \rho }+g_{\beta \gamma }{\widetilde{M}}_{\alpha \rho
}-g_{\beta \rho }{\widetilde{M}}_{\alpha \gamma });  \label{sit2}
\end{equation}
which has two Casimir invariants \cite{g8} 
\[
I_1={\widetilde{M}}_{\alpha \beta }{\widetilde{M}}^{\alpha \beta }, 
\]
and 
\[
I_2=W_\alpha W^\alpha, 
\]
with 
\[
W_\alpha =\varepsilon _{\alpha \beta \gamma \sigma \rho }{\widetilde{M}}%
^{\beta \gamma }{\widetilde{M}}^{\sigma \rho }; 
\]
$\varepsilon _{\alpha \beta \gamma \sigma \rho }$ is the totally
antisymmetric tensor in five dimensions. Observe that now $\widetilde{M} \in 
\mathcal{S}$ (not $M$) is antisymmetric.

As an example, consider a particular case of rotations plus spatial
translations in $\mathcal{G}$ of the type $\ {\bar x}^\mu =G_{\ \nu}^\mu
x^\nu +a^\mu \ $, with the infinitesimal part of $G_{\ \nu }^\mu $ being
determined by $\mathbf{L}$ and $\mathbf{B}$, generators of a Lie algebra of
the Euclidean group (we can consider full inhomogeneous transformations in $%
\mathcal{G}$, but these are not of much interest here. A more detailed
discussion about this point will appear elsewhere). In this case we get the
algebra

\begin{eqnarray}
\lbrack L_i,L_j] &=&\varepsilon _{ijk}L_k,\,\,\,[L_i,P_j]=\varepsilon
_{ijk}P_k,\,\,\,[L_i,B_j]=\varepsilon _{ijk}B_k,  \nonumber \\
\lbrack B_i,P_4] &=&P_i,\,\,\,[B_i,P_j]=P_5\delta _{ij}.  \label{alg2}
\end{eqnarray}

Finite transformations are provided by 
\begin{eqnarray*}
K_i&=&e^{-v^iB_i},\ \ R_{ij}=e^{\epsilon _{ijk}\theta_{ij}L_k}, \\
T_\mu&=&e^{a^\mu P_\mu},
\end{eqnarray*}
where no sum is implied with repeated indices. Consider a vector in $%
\mathcal{G}$ given by $x=(\mathbf{x},t,\mathbf{x}^2/2t)$, then the
components of the transformed vector, ${\bar x}$, are 
\begin{eqnarray}
{\bar x}^i &=&R_{\ j}^ix^j-v^ix^4+ a^i,  \label{gal1} \\
{\bar x}^4 &=&x^4+a^4,  \label{gal2} \\
{\bar x}^5 &=&x^5-v^i(R_{\ j}^ix^j)+{\frac 12}\mathbf{v}^2x^4+a^5.
\label{gal3}
\end{eqnarray}
Eqs.(\ref{gal1}) and (\ref{gal2}) are just the Galilei transformations, Eqs.(%
\ref{trans1}) and (\ref{trans2}), when $x^4=t,\ a^4=b$.

At this point, it is intersting to observe that the natural representation
of the Galilei Lie algebra is obtained from Eq.(\ref{trans1})--(\ref{trans2}%
) with $P_5=0$. But, $P_5$ is a Casimir invariant (having then a constant
value in the representation). Then, in this covariant context, the usual
central extention of the Galilei Group arises naturally, without any
reference to ray, or unfaithfull, representations.

Another example of this formalism is described by the Lagrangian\cite
{takahashi3} 
\begin{eqnarray*}
\mathcal{L} &=&-\frac{\hbar ^2}{2m}\{\nabla \chi ^{*}\cdot \nabla \chi
-\partial _5\chi ^{*}\partial _4\chi -\partial _4\chi ^{*}\partial _5\chi \\
&&\ +B^{*}(x)(\partial _5+\frac{im}\hbar )\chi +(\partial _5+\frac{im}\hbar
)\chi ^{*}B(x)\},
\end{eqnarray*}
where $B(x)$ is an auxiliary field. Following Ref. \cite{takahashi3}, the
scalar Schr\"odinger equation is derived, that is 
\begin{equation}
\partial _\mu \partial ^\mu \chi (x)=0  \label{schr}
\end{equation}
and 
\[
(\partial _5+\frac{im}\hbar )\chi (x)=0. 
\]
Then, $\chi (x)=\exp (-imx^5/\hbar )\psi (\mathbf{x},x^4)$. Since $x^4=t$,
we have from Eq.(\ref{schr})that 
\[
i\hbar \partial _t\psi (\mathbf{x},t)=-\frac{\hbar ^2}{2m}\nabla ^2\psi (%
\mathbf{x},t). 
\]
The energy-momentum tensor is thus 
\[
T_{\,\,\beta }^\alpha (x)=-\frac{\partial \mathcal{L}}{\partial (\partial
_\alpha \chi )}\partial _\beta \chi +\mathcal{L}\delta _{\,\,\beta }^\alpha
. 
\]
Then, the dynamical variables for space translation, $P^i$, time
translation, $H$, mass, $M$, space rotation, $L^i$, and Galilei boost, $B^i$
are given by 
\begin{eqnarray*}
P^i &=&\int d^3x\,dx^5T_{\,\,i}^4, \\
H &=&\int d^3x\,dx^5T_{\,\,4}^4, \\
M &=&\int d^3x\,dx^5T_{\,\,5}^4, \\
L^i &=&\frac 12\varepsilon _{ijk}\int
d^3x\,dx^5(x^jT_{\,\,k}^4-x^kT_{\,\,j}^4), \\
B^i &=&\int d^3x\,dx^5(tT_{\,\,i}^4+x^iT_{\,\,5}^4),
\end{eqnarray*}
with 
\begin{eqnarray*}
T_{\,\,i}^4 &=&\frac{i\hbar }2(\chi ^{*}\partial _i\chi -\partial _i\chi
^{*}\chi ), \\
T_{\,\,4}^4 &=&\frac{\hbar ^2}{2m}\nabla \chi ^{*}\cdot \nabla \chi , \\
T_{\,\,5}^4 &=&m\,\chi ^{*}\chi .
\end{eqnarray*}

Using the commutation relation between the fields as given in Ref.\cite
{takahashi3}, it is easy to show that the operators $P^i$, $H$, $M$, $L^i$
and $B^i$ define a representation for the algebra given by Eq.(\ref{alg2})

Last but not least, we would like to point out that the tensor analysis in $%
\mathcal{G}$ follows the same way as it is usually done \cite{bishop}. First,

consider $w=w_\mu e^\mu $ and $v=v_\mu e^\mu ;\ w,v\in \mathcal{G}^{*}$. The
tensor product of two arbitrary vectors $x$ and $y$ in $\mathcal{G}$ is a
bilinear form defined by the mapping $\ \tau ={x\otimes y}:\mathcal{G}%
^{*}\times \mathcal{G} ^{*}\mapsto \mathcal{R}$, with

\begin{equation}  \label{tensor1}
\tau(w,v)={x\otimes y}(w,v)=w(x)v(x).
\end{equation}
In terms of components, Eq.(\ref{tensor1}) can be written as 
\begin{equation}  \label{tensor2}
{x\otimes y}(w,v)=w_{\mu}v_{\nu}x^{\mu}y^{\nu},
\end{equation}
and the metric $\eta_{\mu\nu}$ can be given by $\eta_{\mu\nu}=e_{\mu}\otimes
e_{\nu}. $ This is so, since we consider the metric as the mapping $\
\eta_{\mu\nu}: \mathcal{G}^{*}\times\mathcal{G}^{*}\mapsto \mathcal{R}$,
such that $\ \eta_{\mu\nu}:(w,v)\mapsto w_{\mu}v_{\nu}$. Then, it follows
from Eq.(\ref{tensor2}) that

\begin{equation}
{x\otimes y}=x^\mu y^\nu e_\mu \otimes e_\nu .  \label{tensor3}
\end{equation}
Using Eq.(\ref{tensor1}), we can show that $\tau ^{\mu \nu }\equiv \tau
(e^\mu ,e^\nu )=x^\mu y^\nu $; as a consequence, $\tau =\tau ^{\mu \nu }\eta
_{\mu \nu }$. The set $\left\{ \eta _{\mu \nu }=e_\mu \otimes e_\nu \right\} 
$ is a basis spanning the vector space defined by the set of 2nd order
contravariant tensors; the proof and the generalization to higher order
tensors are straightforward.

In summary, through an immersion of the Euclidian space in a $(4,1)$-de
Sitter space, we show how to derive a manifold that leads to a covariant
structure of the Galilei symmetries. For instance, for the Euclidian space
of positions, time can be identified as an imbedding parameter, or, in other
words, the classical space-time, $\mathcal{R}^3\times \mathcal{R}$, is
embedded into $(4,1)$-de Sitter space. This realizes the natural
representation of the Galilei group within the defining representation of
the de Sitter Group. We have studied, therefore, a covariant Galilei Lie
algebra and developed the manifold analysis. As an example, the structure of
the scalar field is considerd, resulting in the scalar Schr\"odinger
equation. A more detailed discussion of the connection established here
between the Galilei symmetries and the de Sitter geometric spaces is in
preparation. 
\[
\]
\textbf{Acknowledgments}: The authors would like to thank D. Page and T.
Kopf for their interest in this work and for suggestions, and R. Jackiw and
H. P. K\"unzle for the copy of their papers. This work was supported by the
Natural Sciences and Engineering Research Council of Canada, and the CNPq (a
Brazilian Agency for Research).

\end{document}